\documentclass[lettersize,journal]{IEEEtran}
\usepackage{amsmath,amsfonts}
\usepackage{physics,amsmath}
\usepackage{algorithm}
\usepackage{algpseudocode}
\usepackage{array}
\usepackage{bm}
\usepackage{subcaption}
\usepackage{mathtools}
\usepackage{xcolor}

\usepackage{textcomp}
\usepackage{stfloats}
\usepackage{lipsum}  
\usepackage{url}
\usepackage{verbatim}
\usepackage{graphicx}
\usepackage{cite}
\usepackage{pdfpages}
\usepackage{tikz}
\usetikzlibrary{3d}
\usetikzlibrary{arrows.meta}

\DeclareMathOperator{\EX}{\mathbb{E}}
\newcommand{\matr}[1]{\mathbf{#1}}
\newcommand{\vect}[1]{\mathbf{#1}}
\DeclareMathOperator{\erfc}{erfc}

\begin{document}

\title{
Frequency-Domain Detection for Molecular Communication with Cross-Reactive Receptors}
\author{Meltem Civas,~\IEEEmembership{Student Member,~IEEE},
        Murat Kuscu,~\IEEEmembership{Member,~IEEE},
        and Ozgur B. Akan,~\IEEEmembership{Fellow,~IEEE}
       \thanks{The authors are with the Department of Electrical and Electronics Engineering, Koç University, Istanbul, Turkey (e-mail: \{mcivas16, mkuscu, akan\}@ku.edu.tr).}
      \thanks{Ozgur B. Akan is also with the Internet of Everything (IoE) Group, Electrical Engineering Division, Department of Engineering, University of Cambridge, Cambridge, CB3 0FA, UK (email: oba21@cam.ac.uk).} 
	   \thanks{This work was supported in part by the AXA Research Fund (AXA Chair for Internet of Everything at Ko\c{c} University), The Scientific and Technological Research Council of Turkey (TUBITAK) under Grant \#120E301, European
Union’s Horizon 2020 Research and Innovation Programme through the Marie Skłodowska-Curie Individual Fellowship under Grant Agreement \#101028935, and Huawei Graduate Research Scholarship.}
\thanks{A preliminary version of this work was presented at IEEE International Conference on Communications (ICC) on June 2023.}
}

\maketitle

\begin{abstract}
Molecular Communications~(MC) is a bio-inspired communication paradigm that uses molecules as information carriers, requiring unconventional transceivers and modulation/detection techniques. Practical MC receivers~(MC-Rxs) can be implemented using field-effect transistor biosensor~(bioFET) architectures, where surface receptors reversibly react with ligands. The time-varying concentration of ligand-bound receptors is translated into electrical signals via field effect, which is used to decode the transmitted information. However, ligand-receptor interactions do not provide an ideal molecular selectivity, as similar ligand types, i.e., interferers, co-existing in the MC channel, can interact with the same type of receptors. Overcoming this molecular cross-talk in the time domain can be challenging, especially when Rx has no knowledge of the interferer statistics or operates near saturation. Therefore, we propose a frequency-domain detection~(FDD) technique for bioFET-based MC-Rxs that exploits the difference in binding reaction rates of different ligand types reflected in the power spectrum of the ligand-receptor binding noise. We derive the bit error probability~(BEP) of the FDD technique and demonstrate its effectiveness in decoding transmitted concentration signals under stochastic molecular interference compared to a widely used time-domain detection~(TDD) technique. We then verified the analytical performance bounds of the FDD through a particle-based spatial stochastic simulator simulating reactions on the MC-Rx in microfluidic channels. 
\end{abstract}


\section{Introduction}
Using molecules to encode and transfer information, i.e., Molecular Communications (MC), is nature's way of connecting \emph{bio things}, such as natural cells, with each other. Engineering this unconventional communication paradigm extend our connectivity to synthetic \emph{bio-nano things}, such as nanobiosensors, artificial cells, is the vision that gave rise to the Internet of Bio-Nano Things~(IoBNT). IoBNT is a novel networking framework with potential to enable groundbreaking healthcare and environmental applications~\cite{akan2016fundamentals, akyildiz2020panacea,koca2021molecular,martins2022microfluidic}. 

MC is fundamentally different from conventional electromagnetic communication techniques as it requires novel transceiver architectures as well as new modulation, coding, and detection techniques that can cope with the highly time-varying, nonlinear, and complex channel characteristics in biochemical environments~\cite{kuscu2019transmitter}. The design of MC receivers~(MC-Rxs) and detection techniques has unquestionably attracted the most attention in the literature. However, due to the simplicity it provides in modeling, many of the previous studies considered passive Rx architectures, that are physically unlinked from the MC channel, and thus, of little practical relevance \cite{kuscu2019transmitter}. An emerging trend in MC is to model and design more practical MC-Rxs that employ ligand receptors on their surface as selective biorecognition units, resembling the sensing and communication interface of natural cells. One such design, which was practically implemented in \cite{kuscu2021fabrication}, is based on field-effect transistor biosensors (bioFETs). In this design, ligand-receptor (LR) interactions are translated into electrical signals via field-effect for the decoding of the transmitted information. 

LR interactions are fundamental to the sensing and communication of natural cells. However, the selectivity of biological receptors against their target ligands is not ideal, and this so-called receptor promiscuity results in cross-talk of other types of molecules co-existing in the biochemical environment \cite{mora2015physical,koca2022narrow}. This cross-talk is often dealt with by natural cells through intracellular chemical reaction networks and multi-state receptor mechanisms, such as kinetic proofreading~\cite{kuscu2019channel,koca2022channel}. The same molecular interference problem also applies to abiotic MC-Rxs using ligand receptors. However, the absence of such molecular mechanisms to mitigate interference makes imperative to develop reliable detection techniques for these systems.

In our previous studies on biosynthetic MC-Rxs, we addressed the molecular interference issue by developing time-domain detection techniques leveraging different ligand-receptor binding statistics \cite{kuscu2022detection} and channel sensing methods enabling simultaneous sensing of multiple ligand types, thereby allowing the mitigation of interference~\cite{kuscu2019channel}. However, the techniques developed for biosynthetic MC-Rxs are not  applicable to biosensor-based MC-Rxs as they rely on samples from the bound time intervals of individual receptors to differentiate between interferer and information molecules. In biosensor-based MC-Rxs, the time trajectories of individual receptor states are not accessible, as the received signal is based on the concentration of bound ligands, which is transduced into electrical signals with additional noise. The difficulty in decoding information from time-varying concentrations of bound ligands is exacerbated, especially when the MC-Rx has no knowledge of the statistics of the interference concentration, and it operates near saturation. Therefore, new approaches are required to address the molecular interference issue for biosensor-based MC-Rxs.

In this paper, we introduce a novel frequency-domain detection (FDD) technique designed to address the  challenge of distinguishing between different types of ligands for biosensor-based MC-Rxs. The FDD approach utilizes the power spectral density (PSD) of receptor occupancy fluctuations, i.e., binding noise, which contains the distinct characteristics of the different LR interactions. By leveraging this information, the FDD method enables the estimation of individual ligand concentrations in the channel.

Stochastic and reversible LR interactions can be represented through a one-step binding/unbinding model, where the state transition rates are determined by the binding and unbinding rates for the LR pair \cite{mele2020general}.
Although various types of ligands can interact with the same kind of receptors, these interactions often have different binding and unbinding rates. These differences in reaction rates are manifested in distinct characteristic frequencies \( f_c \), which are also dependent on the concentrations of the involved ligands. The characteristic frequency of the LR pair appears as a cut-off frequency in the Lorentzian-shaped PSD of the binding noise. When multiple types of ligands interact with the receptors, the PSD can be represented as the superposition of the Lorentzian-shaped PSDs. In such a scenario, the characteristic frequencies that define the shape of the binding noise PSD contain information on both reaction rates and the concentrations of the different ligands. The proposed FDD method exploits this feature to estimate the concentration of information-carrying molecules. This estimation is realized with a quasi-maximum likelihood approach, specifically the Whittle likelihood, which is based on both the observed PSD and the model PSD of the electrical output fluctuations at the MC-Rx. This electrical signal incorporates the binding noise from which the concentration information is extracted. The FDD method optimally decodes the transmitted symbol based on the estimated information-carrying molecule concentration.

To our knowledge, the FDD method introduced in this work is a novel approach for biosensor-based MC-Rxs as it addresses the limitation of the existing TDD methods in distinguishing between information-carrying and interfering ligands. To quantify the performance of the FDD method, we derived bit error probability (BEP) in closed form and compared it with the BEP of the TDD method that uses a single sample from the electrical output of the MC-Rx for symbol detection. To validate the accuracy of our analytical results, we used a particle-based stochastic simulator, which allowed us to simulate LR interactions on the surface of the MC-Rx in a microfluidic channel, along with Monte Carlo simulations. Although FDD requires additional computational resources compared to TDD, it is a promising detection strategy for biosensor-based MC-Rxs, especially in high-interference cases where TDD methods underperform significantly. 
In an earlier version of this work~\cite{civas2023frequency}, we derived theoretical performance bounds, while in the current version, we validate the theoretical results using particle-based stochastic simulations and Monte Carlo experiments.

The rest of the paper is organized as follows. Section~\ref{sec:system_mod} provides an overview of the system model, while Section \ref{sec:TDD} details the TDD method. Next, Section~\ref{sec:FDD} introduces the FDD method. Section~\ref{sec:simulation} outlines the validation procedure used to verify the theoretical derivations presented earlier. Following that, Section \ref{sec:eval} offers a comparative evaluation of the BEP performances of TDD and FDD methods. Finally, Section \ref{sec:con} summarizes our conclusions.

\section{System Model}
\label{sec:system_mod}

\begin{figure}
\centering
\resizebox{\columnwidth}{!}{%
\begin{tikzpicture}[font = \huge,font=\sffamily]
  \draw[thick,->] (0,0,0) -- (22,0,0) node[anchor=north east]{\huge\(x\)};
  \draw[thick,->] (0,0,0) -- (0,5,0) node[anchor=north west]{\huge\(z\)};
  \draw[thick,->] (0,0,0) -- (0,0,-7.3) node[anchor=south]{\huge\(y\)};

\node[font = {\Huge\bfseries\sffamily},align=left] at (0,6) {(a)};
\node[font = {\Huge\bfseries\sffamily},align=left] at (0,-3) {(b)};

  \draw[fill=blue,opacity=0.4] (0,0,0) -- (0,3.5,0) -- (0,3.5,-6) -- (0,0,-6) -- cycle;
     \node at (0, 1.5, -3)[rotate=0] {\huge  MC-Tx};  
  \draw (20,0,0) -- (20,3.5,0) -- (20,3.5,-6);
  \draw (0,3.5,0) -- (20,3.5,0) -- (20,3.5,-6);
  \draw (0,0,-6) -- (20,0,-6) -- (20,3.5,-6);
  \draw (0,3.5,-6) -- (20,3.5,-6);
  \draw (20,0,-6) -- (20,0,0);

  \draw[thick,blue,loosely dashed, -{Triangle[open, scale=2]}, bend left] (17,0) to (17,-3.5);
     \node at (11,-8.5,0) {\includegraphics[scale = 0.35,]{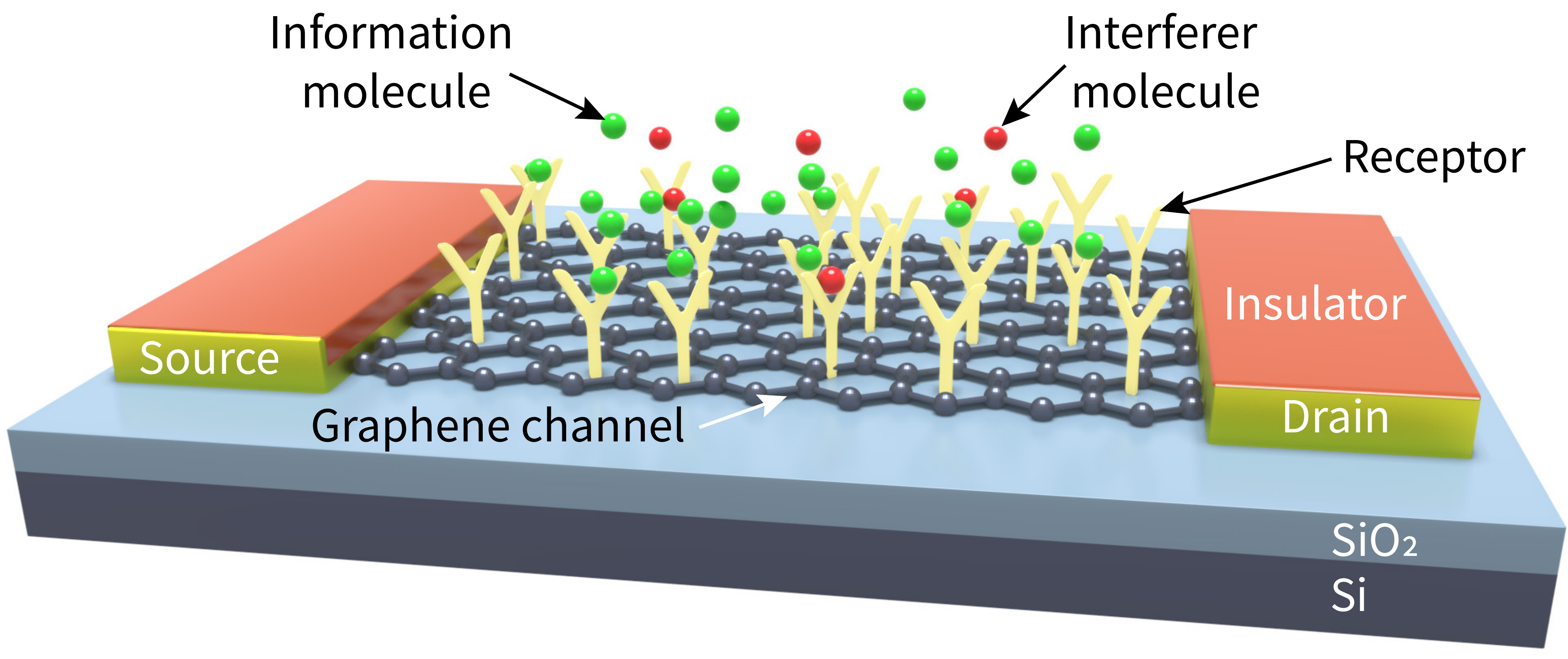}};
 
  \draw (0,0,0) -- (20,0,0) -- (20,0,-6) -- (0,0,-6) -- cycle;

  \draw[fill=blue,opacity=0.4] (15,0,0) -- (15,0,-6) -- (12,0,-6)-- (12,0,0)--cycle;
   \node at (13.5, 0, -3) {\huge MC-Rx};  
 \node at (14.2, -0.42, 1) {\huge \(x =x_R\)};
  \draw[-,thick] (14, 0, 1) -- (14, 0.35, 1);
\draw[rotate=30,thick,blue,loosely dashed](17,-3,10) circle(3cm and 2cm);

  \draw[<->,thick] (8, 0, -6) -- (8, 0, 0) node[midway, left,xshift=-3mm] {\Huge\( w_{\text{ch}} \) };
  \draw[<->, thick] (6, 0, 0) -- (6, 3.5, 0) node[midway, left,yshift=-2mm,xshift=-3mm] {\Huge\( h_{\text{ch}} \) };
\draw[->, thick, red] (19.5, 2.5, -2) -- (20.5, 2.5, -2) node[anchor=west] {\Huge Flow} ;

\end{tikzpicture}}
        \caption{A three-dimensional illustration (a) the microfluidic MC channel, including the positions of the MC-Tx and MC-Rx, and (b) the detailed view of the graphene bioFET-based MC-Rx exposed to information and interferer molecules.}
        \label{gfet}
\end{figure}
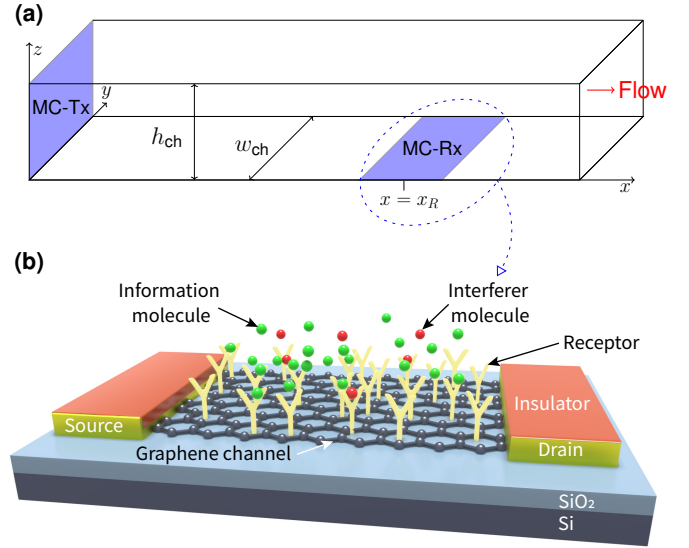

We consider an MC system that employs binary concentration shift keying (CSK). In this scenario, the transmitter (MC-Tx) emits $N_{m|s}$ information molecules at the beginning of each signaling interval. Here, $m$ signifies the information-carrying molecules, while \( s \), which can be either 0 or 1, denotes the transmitted bit. We assume that the signaling interval is long enough to neglect the effect of inter-symbol interference (ISI).
The medium is a three-dimensional microfluidic channel with a rectangular cross-section, as illustrated in Fig.~\ref{gfet}(a). The MC-Tx is located at the channel inlet, while the MC-Rx is positioned on the channel floor, centered at \( x = x_R \). Molecules are released from the MC-Tx instantly and uniformly across the channel cross-section at the release time and propagate through a unidirectional fluid flow along the \( x \)-axis.

The MC-Rx is based on the graphene bioFET architecture, which uses graphene funtionalized with uniform receptors as the transducer channel, as shown in Fig. \ref{gfet}(b). Graphene's unique electrical, mechanical, and biochemical properties, such as high charge carrier mobility, atomic thickness, and two-dimensional geometry combined with the bioFET configuration allowing signal amplification, enable highly sensitive detection of a wide range of biomolecules~\cite{civas2023graphene}. Graphene also provides label-free detection as it allows the functionalization of its surface with biological probes that reversibly bind to the corresponding target. Recently, the first micro/nanoscale MC-Rx based on the graphene bioFET was demonstrated, using single-stranded DNA molecules as information carriers and their complementary strands as receptors on the graphene surface~\cite{kuscu2021fabrication}.

Information molecules can reversibly react with the receptors on the MC-Rx. Additionally, interferer molecules of a single type are present in the channel, and they can react with the MC-Rx receptors as well, albeit with differing reaction rates compared to the information molecules.
The concentration of interferer molecules in the vicinity of the MC-Rx at the sampling time, denoted as \( c_i \), is assumed to follow a log-normal distribution with a mean \( \mu_{c_i} \) and variance \( \sigma^2_{c_i} \). This statistical model is commonly employed for capturing the variability in biological systems \cite{limpert2001log}.

Molecular propagation in the channel is governed by advection-diffusion processes. Given the instantaneous release of \( N_{m|s} \) molecules from the MC-Tx, and under the assumption that receiver reactions negligibly affect the ligand concentration in the channel, the ligand concentration in the channel across space and time can be expressed as follows~\cite{kuscu2018modeling}:
\begin{align}
    c_{m|s}(x,t) = \frac{N_{m|s}}{A_{ch}
\sqrt{4\pi D t}} \exp(-\frac{(x-ut)^2}{4 D t}). 
    \label{c_s}
\end{align}
In~\eqref{c_s}, \( A_{ch} = h_{ch} w_{ch} \) denotes the channel cross-sectional area, \( u \) represents the constant fluid flow velocity along the x-axis, and \( D \) is the effective diffusion coefficient. For channels having a rectangular cross-section, the effective diffusion coefficient \( D \) is formulated in relation to the intrinsic ligand diffusion coefficient \( D_0 \), as follows~\cite{bicen2013system}:
\begin{equation}
    D= \Bigg(1+\frac{8.5 u^2 h_{ch}^2 l_{ch}^2}{210 D_{0}^2(h_{ch}^2+2.4h_{ch}l_{ch}+l_{ch}^2)}\Bigg)D_0.
\end{equation}

The peak of the ligand concentration profile reaches the MC-Rx's center position \( x_R \) at time \( t_D = \frac{x_R}{u} \). Due to the low-pass filtering characteristics of the MC channel induced by the advection diffusion, the ligand concentration at the Rx position varies slowly. This allows the LR reactions to quickly reach equilibrium within a narrow time window centered around \( t_D \) \cite{bicen2013system, kuscu2018modeling}. Within this temporal window, the MC-Rx can sample the receptor states, and the ligand concentration can be expressed as:
\begin{equation}
    c_{m|s}(x_R,t_D)= \frac{N_{m|s}}{A_{ch}\sqrt{4\pi Dt_D}}.
    \label{eqn:peak_con}
\end{equation}
Therefore, the probability of a single receptor being in a bound state in the presence of two different ligand types—information molecules and interferer molecules—can be expressed by 
\begin{equation}
       p_{b|s}= \frac{c_{m|s}/K_{D_m} + c_{i}/K_{D_i}}{1 + c_{m|s}/K_{D_m} + c_{i}/K_{D_i}}, 
   \label{eq:Pb}
\end{equation}
where $K_{D_m} = k^{-}_m/k^{+}_m$ and $K_{D_i} = k^{-}_i/k^{+}_i$ are the dissociation constant of information and interferer molecules, respectively~\cite{kuscu2019channel}.
The total number of receptors in the bound state $N_{b|s}$ follows a Binomial distribution with the mean \( \mu_{N_{b|s}}= p_{b|s} N_r \) and variance \( \sigma_{N_{b|s}}^2 = p_{b|s} (1-p_{b|s}) N_r \) \cite{kuscu2016modeling}. Here, \( N_r \) represents the total number of independent surface receptors.

The binding of charged ligands to the receptors results in an effective charge on the graphene channel via electric field-effect. This can be described by \(Q_{Gr|s} = N_{b|s} q_{eff} N_{e^{-}}\). Here, $N_{e^{-}}$ is number of free electrons per ligand molecules. $q_{eff}$ represents the mean effective charge of a single electron in a bound ligand in the presence of ionic screening, namely, Debye screening. This can be expressed as $q_{eff}= q \times \exp(-\frac{r}{\lambda_D})$. In this expression, $q$ is the elementary charge, $r$ is effective length of a surface receptor, and $\lambda_D$ is the Debye length. The Debye length, \( \lambda_D \), is expressed by \( \lambda_D = \sqrt{\frac{\epsilon \kappa_B T}{2 N_A q^2 c_{ion}}} \), where \( \epsilon \) represents the permittivity of the medium, \( T \) is the temperature, \( \kappa_B \) stands for Boltzmann's constant, \( N_A \) is Avogadro's constant, and \( c_{ion} \) denotes the ionic concentration of the medium~\cite{kuscu2016modeling}.
Accordingly, the mean surface potential due to the bound ligands can be expressed as 
\begin{equation}
    \Psi_{Gr|s}= \frac{Q_{Gr|s}}{C_{G}},
\end{equation}
where $C_{G}= \left(\frac{1}{C_{Gr}}+\frac{1}{C_Q}\right)^{-1}$ is the total gate capacitance of the bioFET. $C_{Gr}$ is the electrical double-layer capacitance between the graphene and electrolyte channel and is defined by \(C_{Gr}=A_{Gr}\epsilon/\lambda_D\). In this expression, $A_{Gr}$ is the area of the graphene surface exposed to the electrolyte. Additionally, $C_Q$ represents the quantum capacitance and is obtained as \(C_Q = c_q \times A_{Gr}\), where $c_q$ is the quantum capacitance of graphene per unit area~\cite{kuscu2021fabrication}.

The change in the output current resulting from bound molecules at equilibrium can be formulated as \(\Delta I_{b|s} = g \times \Psi_{Gr|s}\), where $g$ is the transconductance of the bioFET. For a large number of bound receptors, the number of bound receptors at the sampling time can be approximated as Gaussian-distributed \cite{kuscu2016modeling}, i.e., \( N_{b|s} \sim \mathcal{N}(\mu_{N_{b|s}}, \sigma^2_{N_{b|s}}) \). Considering the linearity of the transduction process, the change in the output current because of bound molecules can also be approximated as following a Gaussian distribution. The mean and variance of this distribution are \( \mu_{\Delta I_{b|s}} = \zeta \mu_{N_{b|s}} \) and \( \sigma^2_{\Delta I_{b|s}} = \zeta^2 \sigma^2_{N_{b|s}} \), respectively. Here, \( \zeta \) is defined as \( \frac{q_{eff} N_{e^{-}} g}{C_G} \).

In low-dimensional semiconductor materials such as graphene, another factor adding to the variability in the overall output current is \(1/f\) noise. According to the widely-used  charge-noise model for graphene FETs \cite{heller2010charge}, the power spectral density of \(1/f\) noise is described by \(S_{f}(f) = \frac{S_{f_{1Hz}}}{f^\beta}, \)
where \(S_{f_{1Hz}}\) is the noise power at 1 Hz, and \(\beta\) is an empirically determined noise exponent with a typical range of \(0.8 \leq \beta \leq 1.2\). 
As discussed in \cite{kuscu2016modeling}, \(1/f\) noise can be approximated as white noise within physically relevant observation windows. Accordingly, the variance of \(1/f\) noise can be formulated as 
\begin{equation}
 \sigma^2_{f} = \int_{0}^{f_L} S_{f}(f_L) \, \mathrm{d}f + \int_{f_L}^{f_H} S_{f}(f) \, \mathrm{d}f, 
 \label{eqn:var_1_f}
\end{equation}
where \(f_L\) represents the lower frequency limit of the observation window, below which the noise power is considered constant, and \(f_H\) is the upper frequency limit, beyond which the noise power is considered negligible. 
As a result, the respective variance and mean of the total output current change are given by \(
     \sigma^2_{\Delta I_s} = \zeta^2 \sigma^2_{N_{b|s}} + \sigma^2_{f}, \mu_{\Delta I_s} = \mu_{\Delta I_{b|s}}. \)

\section{Time-Domain Detection}
\label{sec:TDD}
We assume that Rx has the knowledge of the number of information molecules transmitted, $N_{m|s}$, and the binding/unbinding rates of information and interferer molecules. Since Rx has no knowledge of the interferer concentration statistics, it constructs the optimal ML decision threshold for TDD solely based on its knowledge of the received signal statistics corresponding to the transmitted concentration of information molecules~\cite{kuscu2022detection}:
\begin{equation}
\begin{aligned}
&\gamma_{td} = 
\frac{1}{\sigma_{\Delta I_1}^2-\sigma^2_{\Delta I_0}}\bigg(\sigma^2_{\Delta I_1} \mu_{\Delta I_0} - 
\sigma^2_{\Delta I_0} \mu_{\Delta I_1} +  \sigma_{\Delta I_1} \sigma_{\Delta I_0}\\ & \times \sqrt{(\mu_{\Delta I_1} - \mu_{\Delta I_0})^2 + 2(\sigma^2_{\Delta I_1} -\sigma^2_{\Delta I_0}) \ln(\sigma_{\Delta I_1}/\sigma_{\Delta I_0})} \bigg).
\end{aligned}
\label{th_td}
\end{equation}
As Rx does not account for interference statistics in calculating $\gamma_{td}$, it uses the bound state probability corresponding to a single molecule case, namely, $\displaystyle p_{b|s}= \frac{c_{m|s}/K_{D_m} }{1 + c_{m|s}/K_{D_m}}$.

To derive the BEP for TDD, we first obtain the statistics of the receiver output. By applying the law of total expectation, we can express the mean number of bound receptors as follows:
\begin{equation}
       \mu_{N_{b|s}} = \int_{0}^{\infty}  N_r p_{b|s}(c_i) f(c_i)  \mathrm{d}c_i,       
\end{equation}
where  $\displaystyle p_{b|s}(c_i) = \frac{c_{m|s}/K_{D_m} + c_i/K_{D_i}}{1 + c_{m|s}/K_{D_m} + c_i/K_{D_i}}$, and $f(\cdot)$ is the probability density function of log-normal distribution. Hence, the mean output current change due to bound molecules is 
\begin{equation}
    \mu_{\Delta I_s} = \zeta \int_{0}^{\infty}  N_r p_{b|s}(c_i) f(c_i)  \mathrm{d}c_i.
\end{equation}
Similarly, by applying the law of total variance, we obtain the variance of the current change as 
\begin{equation}
\begin{aligned}
  &\sigma^2_{\Delta I_s} = \zeta^2 \bigg(\int_{0}^{\infty} \left(1-p_{b|s}(c_i)\right) p_{b|s}(c_i) N_r f(c_i) \mathrm{d}c_i \\
  & + \int_{0}^{\infty} \left(p_{b|s}(c_i) N_r\right)^2 f(c_i) \mathrm{d}c_i \bigg)  - \mu_{\Delta I_s}^2 + \sigma^2_{f}.
  \end{aligned}
\end{equation}
Therefore, given the decision threshold $\gamma_{td}$, BEP for TDD can be expressed as follows~\cite{kuscu2022detection}:
\begin{equation}
   P^{TDD}_e = \frac{1}{4} \erfc \left(\frac{\gamma_{td} - \mu_{\Delta I_0}}{\sqrt{2 \sigma^2_{\Delta I_0}}} \right)  +
\frac{1}{4}\erfc \left(\frac{\mu_{\Delta I_1} - \gamma_{td}}{\sqrt{2 \sigma^2_{\Delta I_1}}}\right). 
\end{equation}

\section{Frequency-domain Detection}
\label{sec:FDD}
In this section, we introduce the FDD method utilizing the model and observed PSD  of the overall noise process (binding noise $+$ $1/f$ noise of the graphene bioFET-based MC-Rx) to estimate the received concentration of information molecules $c_{m}$, which will be used in symbol decision. Here, the observed PSD is the periodogram of the noise constructed with the time-domain samples. In the sequel, we  describe the model PSD and then introduce the proposed estimation method.  

\subsection{Theoretical Model of Binding Noise PSD}
This section describes the theoretical model of the binding noise PSD for a particular pair of information and interference concentration, namely $\bm{\lambda} = [c_m, c_i]$.
The binding process of receptors can be described by the reaction model with three states, i.e., unbound (R), bound with information molecules (RM) and bound with interferer molecules (RI), with state occupation probabilities $p_R, p_{RM}$ and $p_{RI}$, respectively \cite{mele2020general}:
\begin{equation}
\begin{aligned}
    R + M &\underset{k_m^+}{\stackrel{k_m^-}{\rightleftharpoons}} RM \\
    R + I &\underset{k_i^+}{\stackrel{k_i^-}{\rightleftharpoons}} RI.
\end{aligned} 
\label{reactions}
\end{equation}
 Hence, the chemical master equations are expressed as follows: 
\begin{equation}
\begin{bmatrix}
\dfrac{\mathrm{d}p_{RM}}{\mathrm{d}t}\\
\dfrac{\mathrm{d}p_{RI}}{\mathrm{d}t}\\
\dfrac{\mathrm{d}p_{R}}{\mathrm{d}t}\\
\end{bmatrix} = 
  \begin{bmatrix}
-k_m^- & 0 & k_m^+ c_{m} \\
0 & -k_i^- & k_i^+ c_{i} \\
k_m^- & k_i^- & -k_m^+ c_{m} -  k_i^+ c_{i}
\end{bmatrix}  
\begin{bmatrix}
p_{RM}\\
p_{RI}\\
p_{R}\\
\end{bmatrix} 
\label{eqn3}
\end{equation} 
The matrix containing reaction rates and the concentrations in \eqref{eqn3}, has rank 2 since one state probability can be written in terms of the other two state occupation probabilities as 
  $p_{R} + p_{RM} + p_{RI} = 1. $
Therefore, by setting the left-hand side in \eqref{eqn3} to zero 
the equilibrium probabilities can be obtained as
\begin{equation}
   p_{RM}^0 = 
 \frac{c_{m}/K_{D_m}}{1 + \frac{c_{m}}{K_{D_m}} + \frac{c_i}{K_{D_i}}},
    p_{RI}^0 = 
 \frac{c_i/K_{D_i}}{1 + \frac{c_{m}}{K_{D_m}} + \frac{c_i}{K_{D_i}}}
\end{equation}
and $p_{R}^0 = 1-(p_{RM}^0 + p_{RI}^0).$ In the equilibrium conditions, the state occupation probabilities can be expressed in terms of the equilibrium state probability and the fluctuations around this probability \cite{mele2020general, mucksch2018quantifying} as
\begin{equation}
     p_{j}(t) = p_{j}^{0} + \Delta p_{j}(t), \quad j \in \{RM, RI, R\}.
     \label{eqn_fluc}
\end{equation}

By substituting \eqref{eqn_fluc} into \eqref{eqn3} and applying Taylor's expansion, the state fluctuations can be expressed as follows~\cite{mele2020general}: 
\begin{equation}
    \frac{\mathrm{d}\Delta \vect{p'}(t)}{\mathrm{d}t} = \matr{\Omega}  \Delta \vect{p'}(t).
    \label{eqnstates}
\end{equation}
We define the column vector $\mathbf{p}(t)$ containing the state probabilities as $
\vect{p}(t) = [ p_{RM}(t), p_{RI}(t), p_{R}(t) ]$
and its reduced form as $
\mathbf{p'}(t) = [ p_{RM}(t), p_{RI}(t) ].$
Therefore, the relation between $\Delta \mathbf{p'}(t)$ 
and $\Delta \mathbf{p}(t)$ can be expressed as:
\begin{equation}
   \Delta \mathbf{p}(t) = \mathbf{R} \Delta \mathbf{p'}(t),
   \label{eq:transform}
\end{equation}
where $\mathbf{R}$ is the transformation matrix \cite{mele2020general}.
Also, the matrix $\mathbf{\Omega}$ in \eqref{eqnstates} is defined as follows:
\begin{equation}
\matr{\Omega} = 
    \begin{bmatrix}
    -k_m^+ c_{{m}}-k_m^- & -k_m^+ \\
    -k_i^+ c_{i} & -k_i^+ c_{i} - k_i^-
    \end{bmatrix}.
    \label{state_probs}
\end{equation}
The deviation in the output current of the MC-Rx due to stochastic binding reactions, i.e., $\Delta I_b(t)$, is then obtained as
\begin{equation}
    \Delta I_b(t) =  \frac{q_{eff}~g}{C_G} ~\vect{z}^T \matr{R} \Delta \vect{p'}(t)
    \label{eqncurrent}
\end{equation}
where $\vect{z} = [N_{e^{-}}; N_{e^{-}}; 0]$ is the vector containing the number of elementary charges. 
As $\Delta I_b(t)$ is a stationary process, the theoretical PSD of the binding noise fluctuations can be obtained by setting $t= 0$ as follows~\cite{mele2020general}:
\begin{equation}
  \begin{aligned}
      &S_b(f) = 2~\mathcal{F} \{\EX [\Delta I_b(t) \Delta I_b(t + \tau)]\} \\ 
     &= 2~\mathcal{F} \{\EX [\Delta I_b(0) \Delta I_b(\tau)]\} \\ 
      &=4 N_r \left(\frac{q_{eff} g}{C_G}\right)^2 \vect{z}^{T}\matr{R}
    \Gamma \left(\Re \{(j 2\pi f \matr{I}_{2\times 2} - \matr{\Omega})^{-1}\}\right)^T \matr{R}^T  \vect{z},
\label{mod_PSD} 
\end{aligned}
\end{equation}
where $\mathcal{F}\{\cdot\}$ stands for Fourier transform, $\matr{I}_{2\times 2}$ is the identity matrix, and $\Gamma$ is the matrix containing the expected state probabilities, which is given as follows \cite{mele2020general}: 
\begin{equation}
\Gamma = 
    \begin{bmatrix}
     p_{RM}^0 \left(1 - p_{RM}^0\right) & -p_{RM}^0 p_{RI}^0 \\
    -p_{RM}^0 p_{RI}^0 & p_{RI}^0 \left(1 - p_{RI}^0\right)
    \end{bmatrix}.
\end{equation}
Therefore, the theoretical PSD of the total current noise corresponding to a particular ($c_m, c_i$) pair can be written as 
\begin{equation}
    S(f) = S_b(f) + S_{f}(f).
    \label{spec}
\end{equation}

\subsection{Maximum Likelihood Estimation of PSD Parameters}
\label{MLE}
In the following part, we describe the parameter value extraction, namely the estimation of information and interfering molecule concentrations, $\bm{\lambda} = [c_m, c_i]$, from the noise PSD. The detector uses the estimated information molecule concentration $\hat{c}_m$ for symbol decision, as will be explained in the following section, Sec.~\ref{sec:detection}. Our analysis is based on the following assumptions:
\begin{itemize}
   \item The total noise process, namely the binding fluctuations combined with $1/f$ noise, is stationary, zero-mean with a single-sided spectrum.
   \item 
   The MC-Rx is provided with the model PSD function, which is expressed by \eqref{spec}.
   MC-Rx also has the knowledge of the binding/unbinding rates of information and interferer molecules, and the number of information molecules 
transmitted for bits $s=0$ and $s=1$ as mentioned in Section~\ref{sec:system_mod}. Therefore, MC-Rx estimates the steady 
information and interferer concentrations by taking time samples from the output current $\Delta I_b$ in a sampling window, where we consider a single realization of the interferer concentration $c_i$ following log-normal distribution as mentioned in Section \ref{sec:system_mod}. The DC component of $\Delta I_b$ is discarded to isolate the noise.
\item 
The information and interferer concentrations are considered constant in the sampling window based on the equilibrium assumption discussed in Section \ref{sec:system_mod} \cite{kuscu2016modeling}. 
\item  
 The observed PSD of time domain samples and the parametric model of the PSD expressed by \eqref{spec} is used in the ML estimation of $\bm{\lambda} = [c_{m}, c_i]$.
 It is assumed that the observed PSD is calculated with the periodogram method.  
\end{itemize}

For each transmitted symbol, we have $N$ number of noise samples $\bm{x} = (x_1, x_2, ..., x_N)$ taken with the sampling period of $\Delta t$. Hence, the total duration of sampling per symbol, namely the length of the sampling window, is $T_d = N \Delta t$. Periodogram for the sampled signal can be computed from the Discrete Fourier transform~(DFT) of the samples $\bm{x}$.
 With even $N$, the periodogram values are then expressed as follows:
    $Y_k = \frac{2\Delta t}{N}|X_k|^2$ where $k = 1,...,N/2-1,$ and $|X_k|$ are DFT components of $\bm{x}.$

For a stochastic time series of length $N$, the random variable $W_k = 2\frac{Y_k}{S(f_k)}$ follows chi-squared distribution $\chi^2$ \cite{vaughan2010bayesian}, 
where $S(f_k)$, given by \eqref{spec}, is the true PSD at frequency $f_k$ and 
$f_k = \frac{k}{ N \Delta t}$ and $k = 1,...,N/2-1$.
The $\chi^2$ distribution with two degrees of freedom is in fact the exponential distribution \cite{barret2012maximum}. Therefore, the periodogram values are exponentially distributed about the true PSD with the following probability given the model PSD value at a given frequency:  
\begin{equation}
	p(Y_k|S(f_k)) = \frac{1}{S(f_k)} \mathrm{e}^{-\frac{Y_k}{S(f_k)}},
	\label{eqn_period}
\end{equation}
following that $S(f_k)$ is also expectation value at $f_k$
\cite{barret2012maximum}. Based on \eqref{eqn_period}, 
the likelihood of observing a pair of particular information and interferer concentrations, $\bm{\lambda} = [c_{m}, c_i]$, is
\begin{equation}
\begin{aligned}
\mathcal{L}(\bm{\lambda}) 
	= 
	\prod_{k = 1}^{N/2-1}p(Y_k|S(f_k,\bm{\lambda})) 
	&= \prod_{k = 1}^{N/2-1} \frac{1}{S(f_k,\bm{\lambda})} \mathrm{e}^{-\frac{Y_k}{S(f_k,\bm{\lambda})}},
\end{aligned}
\end{equation}
where $\bm{\lambda} = [c_m,c_{i}]$ is the parameters to be estimated. Here, we use Whittle likelihood, which can be a good approximation to the exact likelihood asymptotically, and also provide computational efficiency, i.e., $O(N\log N)$ computations compared to $O(N^2)$ for exact likelihood \cite{barret2012maximum, sykulski2019debiased}. Accordingly, the quasi-log likelihood can be written as follows:
\begin{equation}
 \ln\mathcal{L}(\bm{\lambda}) = -\sum_{k = 1}^{N/2-1}\bigg(\frac{Y_k}{S(f_k,\bm{\lambda})} + \ln S(f_k,\bm{\lambda})\bigg).
 \label{eqn7}
\end{equation}
ML estimator extracts the value of $\bm{\lambda}$, i.e., $\hat{\bm{\lambda}}$, that maximizes \eqref{eqn7}. Maximizing $\ln\mathcal{L}(\bm{\lambda})$ is equivalent to minimizing $l = -\ln\mathcal{L}(\bm{\lambda})$ \cite{anderson1990modeling}, such that
\begin{equation}
	\hat{\bm{\lambda}} = \arg\min_{\bm{\lambda}}\{l\}.
\label{eqn8}
\end{equation}
The solution to \eqref{eqn8} can be obtained through numerical methods, such as the Newton-Raphson method achieving ML estimation in only a few iterations. The FDD method requires $O(N \log N)$ computations in addition to the numerical solution of \eqref{eqn8}. In the TDD  approach, however, the MC-Rx takes a single time sample and performs a single thresholding operation to determine the received symbol.

\subsection{Symbol Detection}
\label{sec:detection}

The ML estimator described in Sec.~\ref{MLE} is asymptotically unbiased such that $\hat{\bm{\lambda}}$ tends to have multi-normal distribution \cite{toutain1994maximum} with $\EX[\hat{\bm{\lambda}}] = \bm{\lambda}$, 
and the respective variance of the estimated parameters, which is the diagonal elements of inverse Fisher information matrix~(FIM) $\matr{F}(\bm{\lambda})$, 
\begin{equation}
    \sigma^2_{\hat{\lambda_i}} =  (\matr{F}(\bm{\lambda}))^{-1}_{(ii)}, \quad \matr{F}(\bm{\lambda})_{(ij)} = \EX\left(\frac{\partial^2 l}{\partial \lambda_i \partial \lambda_j}\right).
    \label{IFI}
\end{equation}
where the expectation is taken with respect to the probability distribution of the observed spectrum $p(Y_1,Y_2,...,Y_N)$.
Putting $l=-\ln\mathcal{L}(\bm{\lambda})$ into \eqref{IFI}, the FIM can be expanded as:
\begin{equation}
\begin{aligned}
    &\matr{F}_{(ij)} = \\ &\EX \left( \sum_{k = 1}^{N/2-1} \frac{S(f_k)-Y_k}{S^2(f_k)} \frac{\partial^2 S}{\partial\lambda_i \partial \lambda_j} + \frac{2 Y_k-S(f_k)}{S^3(f_k)} \frac{\partial S}{\partial\lambda_i} \frac{\partial S}{\partial\lambda_j}\right).
\end{aligned}
    \label{FI}
\end{equation}
Considering that $S(f)$ is a slowly varying function, calculating individual periodogram values in \eqref{FI} is unnecessary because  periodogram values can be smoothed by summing over frequency such that 
\begin{equation}
    \sum_{n=1}^{N/2-1} Y_k \phi(f_k) \simeq \sum_{n=1}^{N/2-1} S(f_k) \phi(f_k), 
    \label{eqn:smooth}
\end{equation}
for any smooth function $\phi(f_k)$ ~\cite{levin1965power,toutain1994maximum}. 
Based on \eqref{eqn:smooth}, \eqref{FI} can be simplified as~\cite{toutain1994maximum}
\begin{equation}
    \matr{F}_{(ij)} \simeq \sum_{k=1}^{N/2-1} \frac{1}{S^2(f_k)} \frac{\partial S}{\partial \lambda_i}\frac{\partial S}{\partial \lambda_j},
    \label{fisher}
\end{equation}
where the derivatives are taken at the true value of the parameters. \eqref{fisher} can also be approximated to integral \cite{toutain1994maximum, levin1965power}: 

\begin{equation}
    \matr{F}_{(ij)} \simeq \frac{N\Delta t}{2} \int_{0}^{\frac{1}{2\Delta t}} \frac{1}{S^2(f)} \frac{\partial S}{\partial \lambda_i}\frac{\partial S}{\partial \lambda_j} \mathrm{d}f.
    \label{fisher_integral}
\end{equation}
\eqref{fisher_integral} is a good approximation for a large number of samples such that 
periodogram values can be approximated as Gaussian by the central limit theorem \cite{libbrecht1992ultimate}.

 MC-Rx decides the transmitted bit by applying the ML decision rule to the estimated concentration of the information molecule, denoted as $\hat{c}_m$. In the case of FDD, the ML decision threshold is defined as follows:
\begin{equation}
\begin{aligned}
&\gamma_{fd} = 
\frac{1}{ \sigma^2_{\hat{c}_{m|1}} -\sigma^2_{\hat{c}_{m|0}}} 
\bigg(\sigma^2_{\hat{c}_{m|1}} c_{m|0} - 
\sigma^2_{\hat{c}_{m|0}} c_{m|1} +  \sigma_{\hat{c}_{m|1}} \sigma_{\hat{c}_{m|0}}\\ & \times \sqrt{(c_{m|1} - c_{m|0})^2 + 2(\sigma^2_{\hat{c}_{m|1}} -\sigma^2_{\hat{c}_{m|0}}) \ln(\sigma_{\hat{c}_{m|1}}/\sigma_{\hat{c}_{m|0}})} \bigg),
\end{aligned}
\label{FD_threshold}
\end{equation}
where $\sigma^2_{\hat{c}_{m|s}}$ is the variance and $c_{m|s}$ is the expected value of estimated information molecule when the transmitted bit is $s \in \{0,1\}$.

\begin{algorithm}[!t]
\caption{Frequency Domain Detection }\label{alg:cap}
\begin{algorithmic}[1]
\Function{FrequencyDetection}{$Y[k], N, \Delta t, \gamma_{fd}$}
\State Find initial guess: $\bm{\lambda}^0 \gets [c_m^0, c_i^0]$  
\State $f_k \gets \frac{k}{N\Delta t}$ for $k= 1,\dots, N/2-1$
\State Use Newton's method on \Call{Whittle}{$Y[k],f_k,\bm{\lambda}$} with initial guess $\bm{\lambda}^0$ to find optimal $\bm{\lambda}^{*} = [c_m^{*}, c_i^{*}]$.
\State $\hat{c}_m \gets c_m^{*}$ 
\If {$\hat{c}_m > \gamma_{fd}$} 
    \State Estimated bit: $\hat{s} \gets 1$
\Else
    \State Estimated bit: $\hat{s} \gets 0$
\EndIf

\State \textbf{return} $\hat{s}$
\EndFunction
\Function{Whittle}{$Y[k],f_k,\bm{\lambda}$}
\State $l \gets \sum_{k = 1}^{N/2-1}\frac{Y_k}{S(f_k,\bm{\lambda})} + \ln S(f_k, \bm{\lambda})$
\State \textbf{return} $l$
\EndFunction
\end{algorithmic}
\end{algorithm}

The MC-Rx is assumed to know the peak information molecule concentrations for bit-0 and bit-1 at the sampling time, as expressed by the equation:
\begin{equation}
    c_{m|s} = \frac{N_{m|s}}{A_{ch} \sqrt{4\pi D t_d}}.
\end{equation}
As the MC-Rx does not know the interfering molecule concentration, it computes the decision threshold, $\gamma_{fd}$, assuming the absense of interference in the channel. 
Therefore, the computation of ML decision threshold $\gamma_{fd}$ relies on the following PSD model for a single type of molecule, i.e., information molecule:
\begin{equation}
    S(f,c_m) = 4 N_r \zeta^2 \frac{1}{2\pi f + 1/\tau_{m}} p_{b} (1- p_{b})  +  S_{f}(f),
    \label{no_interference_model}
\end{equation}
where $\tau_{m} = 1/(c_{m} k^{+}_{m} + k^{-}_{m})$ and $p_{b} = \displaystyle\frac{c_{m}}{K_{D_m} + c_{m}}.$
Using \eqref{no_interference_model}, and \eqref{fisher} for $\bm{\lambda} = [c_m]$, the variance of estimated information molecule concentration corresponding to the transmitted bit $s \in \{0,1\}$ is written as \cite{kuscu2022detection}
\begin{equation}
    \sigma^2_{\hat{c}_{m|s}} = \frac{1}{\displaystyle \frac{N\Delta t}{2}\int_{0}^{\frac{1}{2\Delta t}} 
    \frac{1}{S^2(f,c_{m|s})} \left(\frac{\partial S}{\partial c_m}\right)^2\Bigr|_{\substack{c_m = c_{m|s}}}\mathrm{d}f}.
    \label{eqn:sigma}
\end{equation}
\eqref{eqn:sigma} does not give the actual asymptotic variances since MC-Rx estimates the value of $c_m$ based on the model PSD given by \eqref{spec}.

Algorithm~\ref{alg:cap} outlines the FDD procedure, providing the necessary functions for solving the estimation problem defined by equation \eqref{eqn8} and determining the transmitted bit at the receiver.
On line 2, the initial concentration values $[c_m^0, c_i^0]$ are obtained by sweeping over sparse concentration values on \Call{Whittle}{$Y[k], f_k, \bm{\lambda}$} and selecting the values that minimize the corresponding function. On line 3, the variable $f_k$ represents the frequencies for which periodogram values are available. On line 4, Newton-Raphson method is used to solve the estimation problem defined in \eqref{eqn8} by using \Call{Whittle}{$Y[k], f_k, \bm{\lambda}$} and the initial concentration values $[c_m^0, c_i^0]$ as the starting points.
 Once the optimal $\bm{\lambda}^*$ is found, lines 6-11 execute the thresholding operation on the estimated information molecule concentration $\hat{c}_m$ to determine the transmitted bit.

\subsection{Asymptotic Bit Error Probability}
To calculate BEP for FDD, we need the actual values of the variance of estimated information molecule concentration corresponding to $s=0\text{ and }s = 1$, i.e., $\sigma^2_{\hat{c}_{m|s}}$. Using the model PSD $S(f,\bm{\lambda})$ given by \eqref{spec}, and \eqref{fisher} with $\bm{\lambda} = [c_m,c_i]$, 
the variance can be expressed as 
\begin{equation}
    \sigma^2_{\hat{c}_{m|s}}=(\matr{F}_s(\bm{\lambda}))_{(11)}^{-1},
    \label{Fs_inv}
\end{equation} where $\matr{F}_s$'s elements are: 
\begin{equation}
    \begin{aligned}
     \matr{F}_{{s}_{(11)}} &= \\ \frac{N\Delta t}{2}&\int_{0}^{\frac{1}{2\Delta t}} \frac{1}{S^2(f,(c_{m|s},\mu_{c_i}))} \left(\frac{\partial S}{\partial c_m}\right)^2 \Bigr|_{\substack{c_m = c_{m|s} \\c_i = \mu_{c_i}}}\mathrm{d}f\\
\matr{F}_{s_{(22)}} &=  \\ \frac{N\Delta t}{2}&\int_{0}^{\frac{1}{2\Delta t}} \frac{1}{S^2(f,(c_{m|s},\mu_{c_i}))} \left(\frac{\partial S}{\partial c_i}\right)^2 \Bigr|_{\substack{c_m = c_{m|s} \\c_i = \mu_{c_i}}}\mathrm{d}f,\\
\matr{F}_{s_{(12),(21)}} =  &\\
\frac{N\Delta t}{2}\int_{0}^{\frac{1}{2\Delta t}}& \frac{1}{S^2(f,(c_{m|s},\mu_{c_i}))} \left(\frac{\partial S}{\partial c_m}\right) \left(\frac{\partial S}{\partial c_i}\right) \Bigr|_{\substack{c_m = c_{m|s} \\c_i = \mu_{c_i}}} \mathrm{d}f.
\end{aligned}  
\label{Fs}
\end{equation}

As a result, BEP for FDD can be written as 
\begin{equation}
P^{FDD}_e = \frac{1}{4} \erfc \left(\frac{\gamma_{fd} - c_{m|0}}{\sqrt{2 \sigma^2_{\hat{c}_{m|0}}}} \right) +
\frac{1}{4}\erfc \left(\frac{c_{m|1} - \gamma_{fd}}{\sqrt{2 \sigma^2_{\hat{c}_{m|1}}}}\right).
\label{eq:bep_fd}
\end{equation}
Here, it should be noted that \eqref{eq:bep_fd} is an asymptotic expression based on the Gaussian distribution assumption in Sec.~\ref{sec:detection}.

\section{Simulation of the System}
\label{sec:simulation}
In this section, we detailed the simulation process used to validate our FDD technique. Specifically, we conducted Monte Carlo simulations to compare the theoretical results obtained by the analytical expressions derived in Sections \ref{sec:TDD} and \ref{sec:FDD} and the results obtained through particle-based stochastic simulations. For the latter, we used Smoldyn~\cite{smoldyn2022}.  
The simulation algorithms were implemented in MATLAB, and parallelized CPU cores were used for reasonable computation times.

\subsection{Particle-based Stochastic Simulations}
\begin{figure*}[t]
    \centering
    \includegraphics[width = \textwidth]{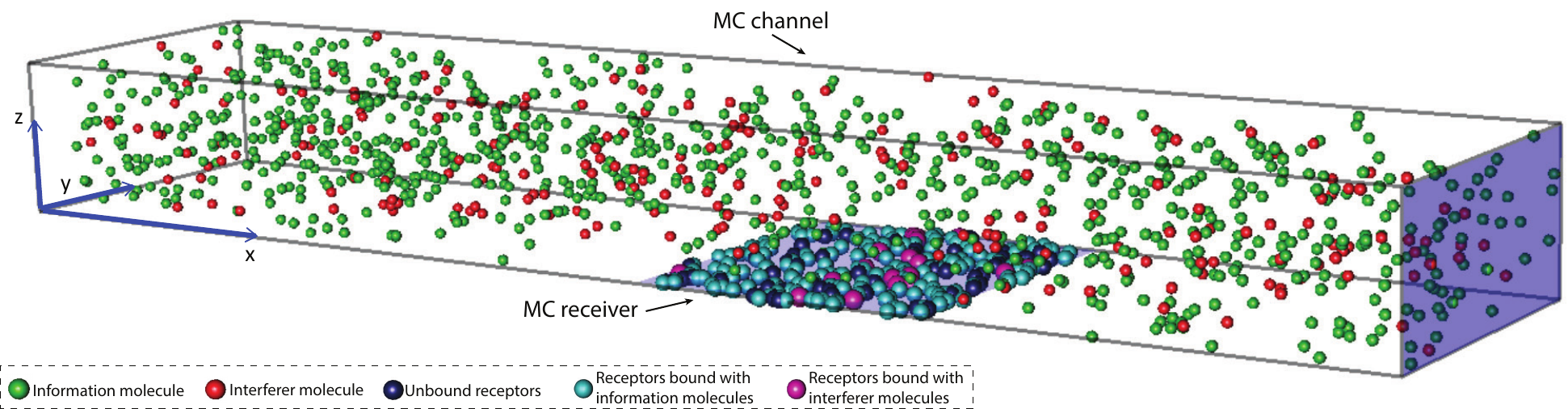}
    \caption{Snapshot of the Smoldyn simulation illustrating the MC system. The parameters have been adjusted for illustrative purposes. Thus, this representation is not to scale.
    }
    \label{fig:smoldyn}
\end{figure*}

We used the open-source particle-based spatial stochastic simulator, \emph{Smoldyn}, to simulate the system. Its capability to capture stochastic events at a molecular level, such as diffusion and receptor binding/unbinding, fits our purpose \cite{smoldyn2022}. As illustrated in Fig. \ref{fig:smoldyn}, we created a straight microfluidic channel with a rectangular cross-section. The receptors are located in a specific region on the channel bed, indicating the position of the MC receiver.

\subsection{Simulation Parameters}

The simulation time step is an important parameter that affects the accuracy of Smoldyn. Using shorter time steps generally leads to higher accuracy but slower simulations. When choosing the time step, one consideration is to ensure it is significantly shorter than the characteristic time scale of any reaction in the system. For the reactions described by \eqref{reactions}, the characteristic times or time constants can be obtained from the matrix $\matr{\Omega}$ defined by \eqref{state_probs}, as follows:
\begin{equation}
\tau_{c_{1,2}} = -1/\lambda_{c_{1,2}},
\label{relaxations}
\end{equation}
where $\lambda_{c_{1,2}}$ corresponds to the eigenvalues of $\matr{\Omega}$. These time constants only depend on the kinetics of the reactions and the concentrations of the information and interferer molecules. Therefore, given the transmitted bit $s$, the characteristic times \eqref{relaxations} can be expanded as 
\begin{equation}
\tau_{c_{1,2|s}} = \frac{2}{\left[\frac{1}{\tau_{m|s}} + \frac{1}{\tau_{i}} \pm \sqrt{\left(\frac{1}{\tau_{m|s}}-\frac{1}{\tau_{i}}\right)^2 + 4 k_m^{+} c_{m|s} k_i^{+}  c_i }\right]},
\end{equation}
where $\tau_{m|s} = 1/(c_{m|s} k_m^+ + k_m^-), \tau_i = 1/(c_i k_i^+ + k_i^-)$ correspond to the characteristic times of the independent reactions  $R + M \underset{k_m^+}{\stackrel{k_m^-}{\rightleftharpoons}} RM $ and $R + I \underset{k_i^+}{\stackrel{k_i^-}{\rightleftharpoons}} RI$ respectively. 

Considering the system parameters summarized in Table \ref{tab:parameters}, we have chosen the simulation time step as $\Delta_{t_s} = 0.0025$ s. This choice represents a good compromise between accuracy and total execution time. Further reducing the time step can significantly increase the simulation time, posing  a challenge for repeated experiments.

\begin{figure}[t]
    \centering
    \begin{subfigure}{0.45\textwidth}
        \includegraphics[width=\linewidth]{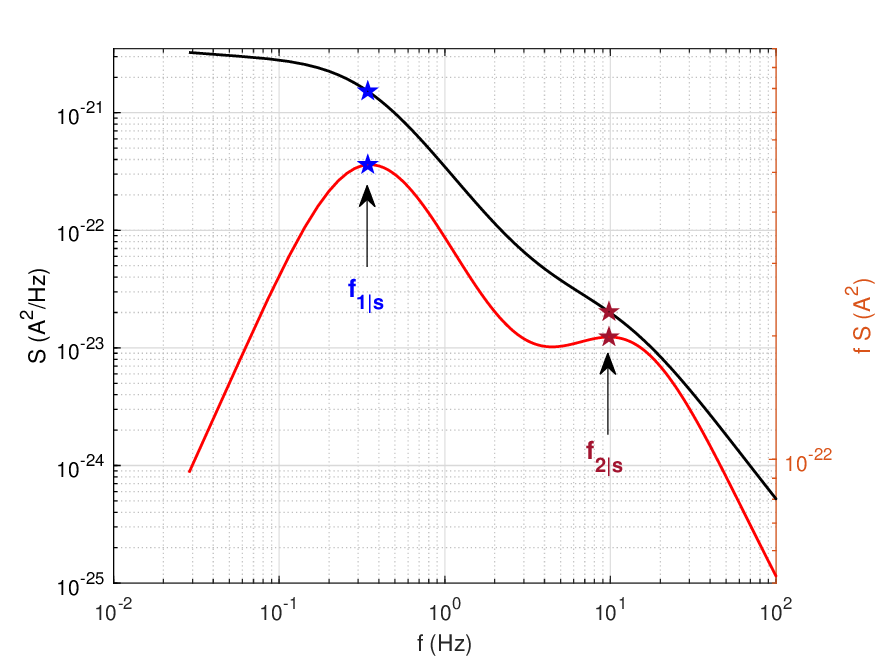}
        \caption{}
        \label{fig:cf}
    \end{subfigure}
    \begin{subfigure}{0.45\textwidth}
        \includegraphics[width=\linewidth]{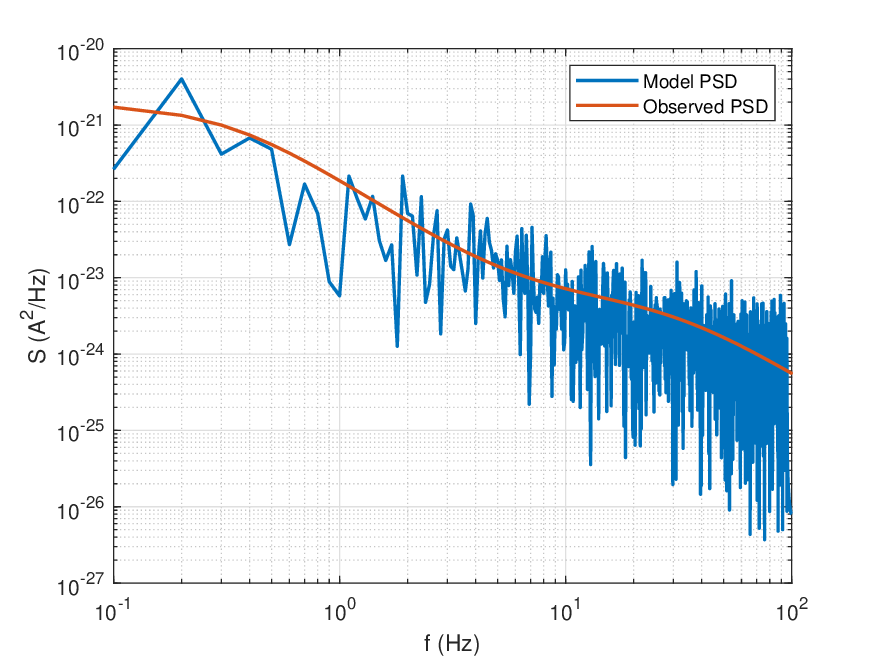}
        \caption{}
        \label{fig:PSD_sim}
    \end{subfigure}
    \caption{ (a) Characteristic frequencies highlighted on the graph of the $S$ multiplied by frequency $f$, where $S$ is the theoretical PSD of noise. (b) Model PSD generated with estimated molecule concentrations and the observed PSD (periodogram) generated with the Smoldyn data.}
    \label{fig:PSD}
\end{figure}

The noise PSD, given by equations \eqref{mod_PSD}-\eqref{spec}, can be described as a superposition of Lorentzian spectral profiles. When they are not masked, the corner frequencies of these Lorentzian profiles are defined by the characteristic frequencies, which can be expressed in terms of characteristic times as $f_{c_{1,2|s}} = 1/(2\pi\tau_{c_{1,2|s}})$. When these frequencies are not masked, they can be identified by locating the peaks of the noise PSD multiplied by the corresponding frequency, as illustrated in Fig. \ref{fig:cf}. Since the PSD exhibits a profile defined by its characteristic frequencies, which contain information about the equilibrium concentration of the information molecules within the sampling window, it is crucial to resolve these characteristic frequencies to extract the relevant information accurately. In this regard, the sampling period $\Delta t$ is a relevant parameter. The maximum observable frequency can be determined by $f_{c_{\text{max}}} = \frac{1}{2\Delta t}$. When a characteristic frequency exceeds this limit, certain portion of the information can be lost.

\begin{algorithm}[!t]
\caption{Monte Carlo Simulation}
\label{alg:MC}
\begin{algorithmic}[1]
\State \textbf{Initialize:} TDD error $e_{t} \gets 0$, FDD error $e_{f} \gets 0$
\State Generate a \textit{bitstream} consisting of $M$ bits
    \State Find decision threshold for TDD $\gamma_{td}$ 
    \State Find decision threshold for FDD $\gamma_{fd}$
\For{each bit $s$ in \textit{bitstream}}
    \State Compute information concentrations $c_{m|s}$ with \eqref{eqn:peak_con}
    \State Generate interferer concentration $c_i \sim \log\mathcal{N}(a,b)$
    \State Run Smoldyn. Wait until equilibrium.
    \State Record $N$ samples from the number of bound receptors $n_B[k]$ for $k = 1, \ldots, N$ with sampling period $\Delta t$
    \State Generate $1/f$ noise samples in the time domain $n_{1/f}[k]$ for $k = 1, \ldots, N$  
    \State Convert to electrical output $I_b[k] \gets n_{1/f}[k] + \zeta n_B[k]$  for $k = 1, \ldots, N$ 
    \State Take one sample, denoted as $I_1$, from $I_b[k]$
    \State $\hat{s_t} \gets \Call{TimeDetection}{\gamma_{td},I_1}$
    \State If $\hat{s_t} \neq s$, increment $e_t$ by 1.
        \State Isolate the noise, denoted as $\Delta I_{b}[k]$, from $I_{b}[k]$
    \State Apply low pass filter to $\Delta I_{b}[k]$ to obtain $\Delta I_{b_f}[k]$
    \State Compute the periodogram $Y[m]$ for $\Delta I_{b_f}[k]$, for $m = 1,\ldots,\frac{N}{2}-1$ and sampling period $\Delta t$
    \State $\hat{s_f} \gets \Call{FrequencyDetection}{Y[m], N, \Delta t, \gamma_{fd}}$
    \State If $\hat{s_f} \neq s$, increment $e_f$ by 1.
\EndFor
\State Compute BEP for TDD: $P^{TDD}_e \gets e_t/M$
\State Compute BEP for FDD: $P^{FDD}_e \gets e_f/M$
\Function{TimeDetection}{$\gamma_{td}, I_1$}
    \If{$I_1 > \gamma_{td}$}
        \State $\text{Estimated bit: } \hat{s} \gets 1$
    \Else
        \State $\text{Estimated bit: } \hat{s} \gets 0$
    \EndIf
    \State \textbf{return} $\hat{s}$ 
\EndFunction
\end{algorithmic}
\end{algorithm}

\subsection{Monte Carlo Simulations}

The Monte Carlo algorithm used to validate the BEP performance of FDD is described in Algorithm \ref{alg:MC}. In each iteration (line 6-19), representing the transmission of a random bit $s$ in the microfluidic channel, the procedure is as follows.

Based on the equilibrium assumption explained in Sec.~\ref{sec:system_mod}, the information molecule concentration $c_{m|s}$ and interferer concentration $c_i$ are assumed to remain constant within the sampling window. Therefore, the corresponding information concentration $c_{m|s}$ for the generated bit $s$ and the interferer concentration $c_i$ generated from a log-normal distribution represent the concentrations at the sampling time. With these initial values, the Smoldyn simulation is run for a time duration sufficient for the binding reactions to reach equilibrium.
Once equilibrium is attained, time samples are taken from the output of the Smoldyn (line 9), where the output corresponds to the number of bound receptors. The collected time samples are then scaled and combined with 1/f noise to simulate the transduction process of the bioFET, resulting in the electrical output denoted as $I_b[k]$. 

To simulate the 1/f noise in the time domain, we employed the FIR filtering method described in \cite{kasdin1995discrete}, which involves filtering white Gaussian noise in the frequency domain using an FIR filter with a 1/f passband and subsequently performing an inverse Fourier transform to obtain the 1/f noise sequence. The approximate noise variance $\sigma^2_{f}$ is computed based on the expression given in Equation \eqref{eqn:var_1_f}, with the parameter values specified in Table \ref{tab:parameters}.

As stated in line 12, a single sample is taken from the middle of the electrical output $I_b[k]$ for TDD. 
For FDD, the base signal is removed from the electrical output $I_b[k]$ to isolate the noise component, $\Delta I_b[k]$. Additionally, considering that samples are taken within a finite sampling window,  distortions near the Nyquist frequency in the PSD are inevitable due to the aliasing. Therefore, a low-pass filter with a passband near the Nyquist frequency is applied to the sampled noise to mitigate this distortion, as outlined in line 16. This filtering step helps reduce outliers that could affect the accuracy of the MLE.
Then, periodogram PSD is computed using an N-point FFT of the filtered noise samples. No further processing is performed on the noise signal or the estimated PSD before the MLE to avoid introducing statistical distortion, which can result in erroneous parameter predictions. Finally, the MLE is performed by MATLAB's \textit{fminunc} solver~\cite{Optimization}, which uses the quasi-Newton algorithm. The solver can find the solution within a few iterations in this specific scenario. Finally, BEP for TDD and FDD are computed as described by line 20-21. 

Fig.~\ref{fig:PSD_sim} depicts the model noise PSD generated using the MLE estimated information and interferer concentration pair, $[\hat{c}_{m|s}, \hat{c}_{i}]$, and the periodogram estimate of the observed noise PSD obtained by using the data from the Smoldyn.

\begin{figure*}[t]
    \centering
    \begin{subfigure}{0.45\textwidth}
        \includegraphics[width=\textwidth]{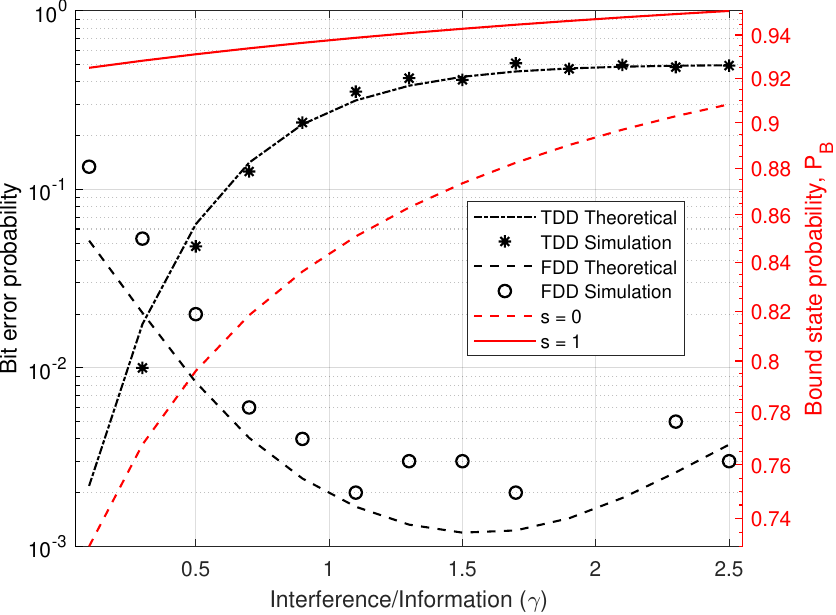}
        \caption{}
        \label{fig:interference}
    \end{subfigure}
    \hspace{0.8cm}  
    \begin{subfigure}{0.45\textwidth}
        \includegraphics[width=\textwidth]{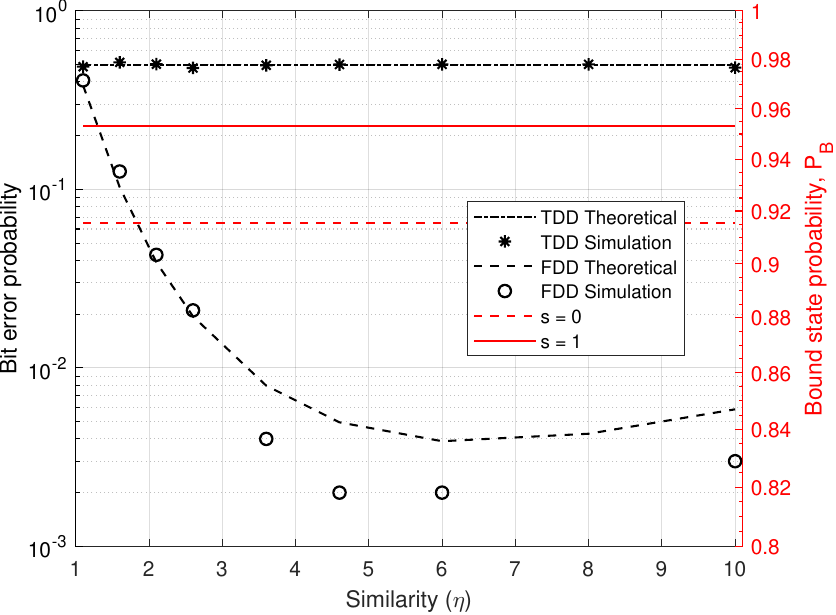}
        \caption{}
        \label{fig:similarity}
    \end{subfigure}
    \caption{BEP for varying (a) information to interference ratio \( \gamma \) and (b) similarity \( \eta \).}
\end{figure*}

\section{Performance Evaluation}
\label{sec:eval}
In this section, we analyze the performance of FDD and TDD in terms of BEP. The default values of the system parameters are given in Table~\ref{tab:parameters}, with the reaction rates adopted from \cite{kuscu2022detection}.  Unless stated otherwise, the default parameter values are used in the rest of the paper. 

It can be observed that the simulation results can deviate from the theoretical results due to the inherent nature of Monte Carlo simulations. Especially when dealing with high concentration values (high signal variance), Monte Carlo simulations can require a large number of iterations to converge, potentially leading to excessive simulation times when using Smoldyn. Therefore, to maintain a compromise between convergence and simulation times, we chose moderate  number of iterations (1000) for the Monte Carlo simulations.

\begin{table}[!b]\scriptsize		
\caption{Default Values of System Parameters}
\label{tab:parameters}
\centering
\begin{tabular}{m{4.9cm}|m{3cm}}
\hline 
 Temperature~($T$) & $300$ K \\ \hline
 Microfluidic channel height ($h_{ch}$), width ($w_{ch}$) & 5 $\mu$m, 10 $\mu$m  \\ \hline
Average flow velocity ($u$) & 10 $\mu$m/s  \\ \hline
 Distance of Rx's center position to Tx ($x_R$) & $1$ mm \\ \hline
 Ionic concentration of medium ($c_{ion}$) & 30 mol/m$^3$  \\ \hline
Relative permittivity of medium ($\epsilon/\epsilon_0$) & $80$  \\ \hline
 Intrinsic diffusion coefficient ($D_0$) & $2 \times 10^{-11}$ m$^2$/s  \\ \hline
Binding rate of information and interferer molecules ($k^{+}_m, k^{+}_i$) & $4 \times 10^{-17}$ m$^3$/s \\ \hline
Unbinding rate of information molecules ($k^{-}_m$) & 2 s$^{-1}$  \\ \hline
Unbinding rate of interferers ($k^{-}_i$) & 8 s$^{-1}$  \\ \hline
Average \# of electrons in a ligand ($N_{e^{-}}$) & 3  \\ \hline
Number of independent receptors ($N_r$) & $120$  \\ \hline
Effective length of a surface receptor ($r$) & 2 nm  \\ \hline
Transconductance of graphene bioFET ($g$)& $1.9044\times 10^{-4}$ A/V \\ \hline
Width of graphene in transistor ($l_{gr}$) & 10$\mu$m  \\ \hline
Quant. capacitance of graphene per unit area 
($c_q$)& $2\times 10^{-2}$ F m$^{-2}$\\ \hline
\# of transmitted ligands for $s = {0,1}$ ($N_{m|s}$) & $[1,5] \times 10^3$  \\ \hline
\# of noise samples ($N$) &  700 \\ \hline
Sampling period ($\Delta t$) & 0.005 s \\ \hline
Mean interference to information concentration ratio ($\gamma = \mu_{c_i}/c_{m|s=1}$) &  0.7 \\ \hline
Interference mean/std ratio ($\mu_{c_i}/\sigma_{c_i}$) &  10 \\ \hline
Power of $1/f$ noise at 1 Hz ($S_{f_{1Hz}}$) & $10^{-23}$ A$^2/$Hz \\ \hline
Noise exponent for 1/f noise ($\beta$) & 1 \\ \hline
Lower frequency limit for 1/f noise ($f_L$) & $10^{-8}$ \\ \hline
Upper frequency limit for 1/f noise ($f_H$) & $10^{7}$ \\ \hline
\end{tabular}%
\end{table}

\subsection{Effect of Interference on BEP}
In the initial analysis, we examine the impact of interference strength on the BEP performance of both TDD and FDD. To investigate this, we introduce a tuning parameter $\gamma$, which determines the mean interferer concentration $\mu_{c_i}$ as $\mu_{c_i} = \gamma \cdot c_{m|s=1}$.
The results, as depicted in Fig.~\ref{fig:interference}, show that FDD outperforms TDD, except in cases where the interferer concentration is significantly lower than the concentration of information molecules as shown in Fig.~\ref{fig:interference}. As the interferer concentration increases, the performance of TDD deteriorates. This degradation occurs because the increasing number of bound receptors becomes occupied by the interfering molecules, leading to saturation of the receiver, thereby causing increase in the error probability.
On the other hand, the performance of FDD improves with increasing interferer concentration up to a certain point, after which it starts to degrade. 
Considering that the characteristic frequencies are influenced by both the concentration and affinities of the molecules, the effect of interference is initially indistinguishable in the PSD. This is because the concentration of interfering molecules is initially lower, which results in the individual Lorentzian PSD attributed to interfering molecules being overshadowed by that of the information-carrying molecules. However, as the concentration of interfering molecules increases, its impact on the combined PSD becomes more pronounced, allowing for a more accurate estimation of concentrations and, consequently, reduced error probabilities. Nevertheless, beyond a particular concentration of interfering molecules, the Lorentzian PSD attributed to the information-carrying molecules is overshadowed, increasing error probabilities.

\begin{figure*}[t]
    \centering
    \begin{subfigure}[b]{0.45\textwidth}
        \includegraphics[width=\textwidth]{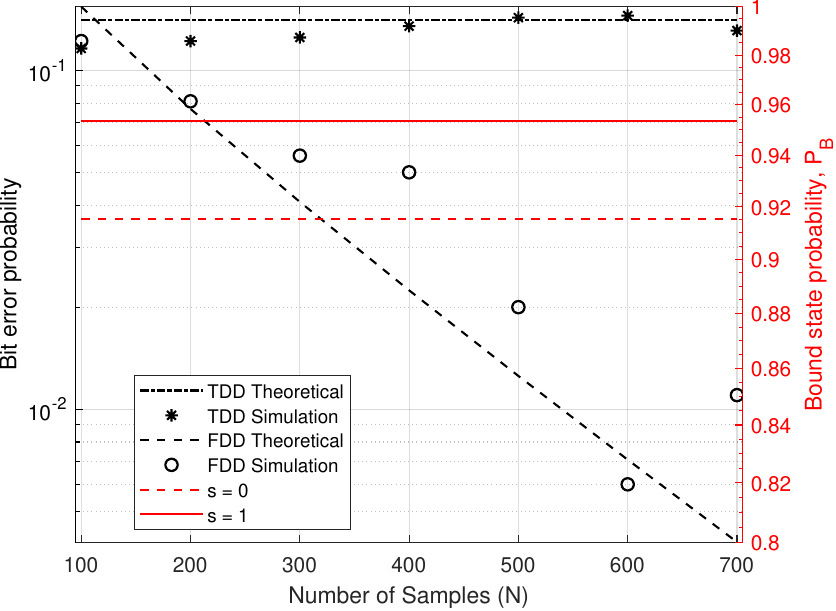}
        \caption{}
        \label{fig:numSamples}
    \end{subfigure}
    \hspace{0.8cm}
    \begin{subfigure}[b]{0.45\textwidth}
        \includegraphics[width=\textwidth]{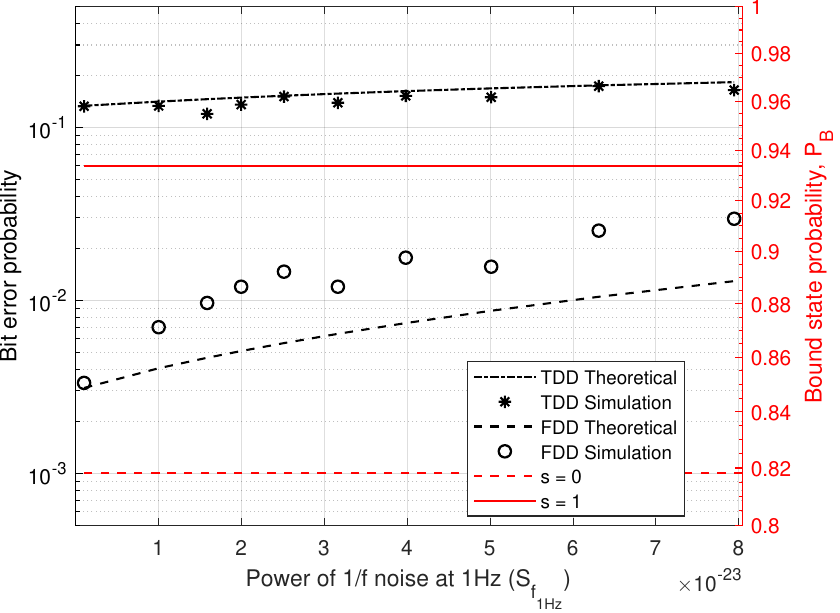}
        \caption{}
        \label{fig:f_noise}
    \end{subfigure}
    \caption{BEP for varying (a) number of samples (b) power of \(1/f\) noise at \(1\) Hz.}
\end{figure*}

\subsection{Effect of Similarity on BEP}
We also consider the effect of the affinity ratio between the information and interferer molecules, which we refer as the similarity parameter and define as $\eta = \frac{K_{D_i}}{K_{D_m}}$.
To observe the direct effect of the similarity, we keep the bound state probabilities constant by scaling the mean interferer concentration, $\mu_{c_i}$, by $\eta$. As expected, TDD is not affected by the similarity parameter, as shown in Figure~\ref{fig:similarity}. 
This occurs because the TDD method relies on the total bound state probability \( p_{b_s} \), which remains constant in this case, thereby not affected by changes in similarity. 
On the other hand, BEP for FDD exhibits an interesting behavior. BEP improves as the similarity parameter $\eta$ increases, indicating that the information and interfering molecules become dissimilar. However, beyond a specific value of the similarity parameter, BEP starts to degrade.
This trend can be explained as follows: When the similarity parameter $\eta$ is low, indicating high similarity between the information and interfering molecules, FDD cannot distinguish between these two types of molecules based on their affinities. This is because any noticeable distinction is not reflected in the spectrum. For example, when the similarity parameter $\eta$ is close to 1, the noise PSD appears as a single Lorentzian profile, thereby obscuring the Lorentzian profile associated with the reaction $RM$ in the spectrum. As the similarity increases, it becomes more visible that PSD is the superposition of two Lorentzian profiles, and the characteristic frequencies $f_{1|s}$ and $f_{2|s}$ are more distinguishable. However, once the similarity goes beyond the point where $f_{1|s}$ and $f_{2|s}$ are most distinguishable, the PSD becomes less informative about the concentration of information $c_{m|s}$. This occurs because interference becomes the dominant factor on the profile of the PSD, and the variations in $c_{m|s}$ have a lesser impact on the PSD. As a result, the PSD becomes less informative about the information concentration, $c_{m|s}$, increasing the error probability.

\subsection{Effect of Number of Samples on BEP}
The BEP performance is further analyzed for the number of time samples $N$. As depicted in Fig.~\ref{fig:numSamples}, it is observed that the performance of FDD improves as the number of samples increases. This outcome aligns with expectations, as a larger number of samples leads to a decrease in the variance of the estimated information molecule concentration, $\sigma^2_{\hat{c}_{m|s}}$. The reduction in variance results in more accurate estimation as the error on the estimated parameter follows $\sigma_{\hat{c}_{m|s}} \propto \sqrt{(\matr{F}_s)_{(11)}^{-1}/N}$. Consequently, more samples lead to a decrease in the BEP.
On the other hand, the performance of TDD remains unaffected by the number of samples, as the receiver only takes one sample, typically from the middle of the signal within the sampling window.

\subsection{Effect of 1/f noise on BEP}
We also simulate the BEP performance of FDD and TDD under varying $1/f$ noise power (at $1$ Hz), as depicted in Fig.~\ref{fig:f_noise}. To better observe the performance trend, a similarity value of $\eta = 3$ is chosen for this case. The results show that 1/f noise affects the performance of both TDD and FDD. Additionally, it can be observed that the performance of FDD is more sensitive to changes in the 1/f noise levels compared to TDD. Since 1/f noise is directly additive across the entire spectrum, increasing the 1/f noise level reduces the FDD performance by potentially masking critical frequencies under the background 1/f noise. 

\begin{figure*}[t]
    \centering
    \begin{subfigure}[b]{0.45\textwidth}
        \includegraphics[width=\textwidth]{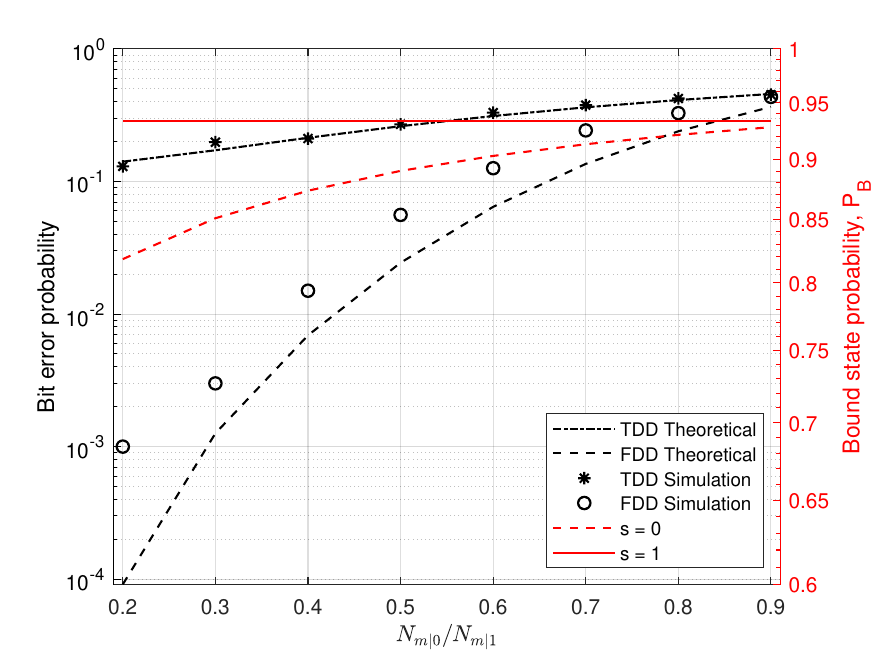}
        \caption{}
        \label{fig:nt1}
    \end{subfigure}
    \hspace{0.8cm} 
    \begin{subfigure}[b]{0.45\textwidth}
        \includegraphics[width=\textwidth]{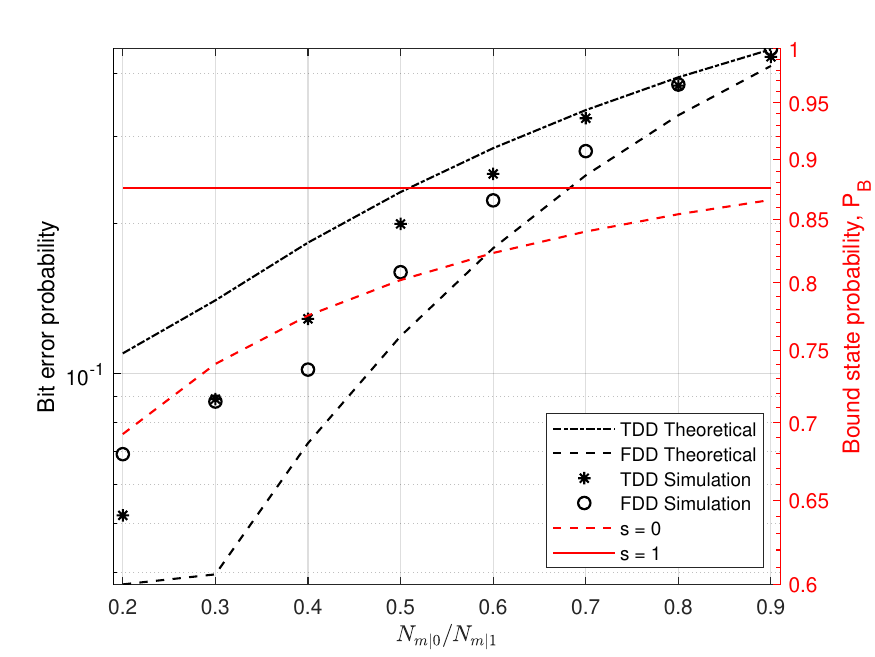}
        \caption{}
        \label{fig:nt2}
    \end{subfigure}
\caption{BEP for varying bit-0/bit-1 concentration ratio when the MC-Rx is (a) near saturation, and (b) in a non-saturation state.}
\end{figure*}

\subsection{Effect of Bit-0/Bit-1 concentration ratio on BEP}
We further explored the impact of the ratio of concentrations for bit-0 and bit-1, denoted as $N_{m|0}/N_{m|1}$, on the BEP. As depicted in Fig. \ref{fig:nt1}, it becomes apparent that the detection performances of both TDD and FDD methods experience a decline when the concentration values become closer. 

In Fig.\ref{fig:nt1}, the MC-Rx is observed in near saturation, implying a high level of receptor occupancy. Comparing the performance in Fig.\ref{fig:nt1} with the result shown in Fig.\ref{fig:nt2}, which corresponds to a case of lower bound state probabilities, a noticeable performance decline can be observed in the latter.

These results suggest that the FDD method is more effective when the receptor occupancy at the MC-Rx is higher. This property could be advantageous, as it aligns more closely with the typical conditions encountered in MC, wherein the receiver is commonly expected to operate near saturation most of the time due to the remaining molecules from previous transmissions or interferers.

\section{Conclusion}
\label{sec:con}

In this paper, we propose a frequency-domain detection method for the bioFET-based molecular communication receivers. The method leverages the receptor cross-reactivity to different ligands and utilizes the output noise power spectral density to detect the transmitted bit. We derived the bit error probability for frequency domain and one-shot time domain detection methods by considering a microfluidic channel where a single type of interfering molecules is present.
We employed a particle-based spatial simulator to validate our theoretical derivations for bit error probabilities. Our analysis indicates a significant performance advantage of the proposed frequency-domain detection method over the time-domain detection method, especially in scenarios with high interference in the channel. 
This characteristic of the frequency-domain detection can be helpful since it can be applicable to practical conditions encountered in molecular communication, where receivers often operate close to saturation at high data transmission rates due to the presence of remaining molecules from previous transmissions or interference. 

\bibliographystyle{IEEEtran}
\bibliography{references}

\begin{thebibliography}{10}
\providecommand{\url}[1]{#1}
\csname url@rmstyle\endcsname
\providecommand{\newblock}{\relax}
\providecommand{\bibinfo}[2]{#2}
\providecommand\BIBentrySTDinterwordspacing{\spaceskip=0pt\relax}
\providecommand\BIBentryALTinterwordstretchfactor{4}
\providecommand\BIBentryALTinterwordspacing{\spaceskip=\fontdimen2\font plus
\BIBentryALTinterwordstretchfactor\fontdimen3\font minus
  \fontdimen4\font\relax}
\providecommand\BIBforeignlanguage[2]{{%
\expandafter\ifx\csname l@#1\endcsname\relax
\typeout{** WARNING: IEEEtran.bst: No hyphenation pattern has been}%
\typeout{** loaded for the language `#1'. Using the pattern for}%
\typeout{** the default language instead.}%
\else
\language=\csname l@#1\endcsname
\fi
#2}}

\bibitem{akan2016fundamentals}
O.~B. Akan, H.~Ramezani, T.~Khan, N.~A. Abbasi, and M.~Kuscu, ``Fundamentals of
  molecular information and communication science,'' \emph{Proc. IEEE}, vol.
  105, no.~2, pp. 306--318, 2016.

\bibitem{akyildiz2020panacea}
I.~F. Akyildiz, M.~Ghovanloo, U.~Guler, T.~Ozkaya-Ahmadov, A.~F. Sarioglu, and
  B.~D. Unluturk, ``Panacea: An internet of bio-nanothings application for
  early detection and mitigation of infectious diseases,'' \emph{IEEE Access},
  vol.~8, pp. 140\,512--140\,523, 2020.

\bibitem{koca2021molecular}
C.~Koca, M.~Civas, S.~M. Sahin, O.~Ergonul, and O.~B. Akan, ``Molecular
  communication theoretical modeling and analysis of sars-cov2 transmission in
  human respiratory system,'' \emph{IEEE Transactions on Molecular, Biological
  and Multi-Scale Communications}, vol.~7, no.~3, pp. 153--164, 2021.

\bibitem{martins2022microfluidic}
D.~P. Martins, M.~T. Barros, B.~J. O’Sullivan, I.~Seymour, A.~O’Riordan,
  L.~Coffey, J.~B. Sweeney, and S.~Balasubramaniam, ``Microfluidic-based
  bacterial molecular computing on a chip,'' \emph{IEEE Sensors Journal},
  vol.~22, no.~17, pp. 16\,772--16\,784, 2022.

\bibitem{kuscu2019transmitter}
M.~Kuscu, E.~Dinc, B.~A. Bilgin, H.~Ramezani, and O.~B. Akan, ``Transmitter and
  receiver architectures for molecular communications: A survey on physical
  design with modulation, coding, and detection techniques,'' \emph{Proceedings
  of the IEEE}, vol. 107, no.~7, pp. 1302--1341, 2019.

\bibitem{kuscu2021fabrication}
M.~Kuscu, H.~Ramezani, E.~Dinc, S.~Akhavan, and O.~B. Akan, ``Fabrication and
  microfluidic analysis of graphene-based molecular communication receiver for
  internet of nano things (iont),'' \emph{Scientific reports}, vol.~11, no.~1,
  pp. 1--20, 2021.

\bibitem{mora2015physical}
T.~Mora, ``Physical limit to concentration sensing amid spurious ligands,''
  \emph{Physical review letters}, vol. 115, no.~3, p. 038102, 2015.

\bibitem{koca2022narrow}
C.~Koca, M.~Civas, and O.~Akan, ``Narrow escape problem in synaptic molecular
  communications,'' in \emph{Proceedings of the 9th ACM International
  Conference on Nanoscale Computing and Communication}, 2022, pp. 1--7.

\bibitem{kuscu2019channel}
M.~Kuscu and O.~B. Akan, ``Channel sensing in molecular communications with
  single type of ligand receptors,'' \emph{IEEE Transactions on
  Communications}, vol.~67, no.~10, pp. 6868--6884, 2019.

\bibitem{koca2022channel}
C.~Koca and O.~B. Akan, ``Channel clearance by perfectly absorbing boundaries
  in synaptic molecular communications,'' \emph{IEEE Access}, vol.~10, pp.
  121\,480--121\,493, 2022.

\bibitem{kuscu2022detection}
M.~Kuscu and O.~B. Akan, ``Detection in molecular communications with ligand
  receptors under molecular interference,'' \emph{Digital Signal Processing},
  vol. 124, p. 103186, 2022.

\bibitem{mele2020general}
L.~J. Mele, P.~Palestri, and L.~Selmi, ``General model and equivalent circuit
  for the chemical noise spectrum associated to surface charge fluctuation in
  potentiometric sensors,'' \emph{IEEE Sensors Journal}, vol.~21, no.~5, pp.
  6258--6269, 2020.

\bibitem{civas2023frequency}
M.~Civas, A.~Abdali, M.~Kuscu, and O.~B. Akan, ``Frequency-domain detection for
  molecular communications,'' \emph{Proceedings of IEEE ICC 2023}, 2023.

\bibitem{civas2023graphene}
M.~Civas, M.~Kuscu, O.~Cetinkaya, B.~E. Ortlek, and O.~B. Akan, ``Graphene and
  related materials for the internet of bio-nano things,'' \emph{APL
  Materials}, vol.~11, no.~8, p. 080901, 2023.

\bibitem{limpert2001log}
E.~Limpert, W.~A. Stahel, and M.~Abbt, ``Log-normal distributions across the
  sciences: keys and clues: on the charms of statistics, and how mechanical
  models resembling gambling machines offer a link to a handy way to
  characterize log-normal distributions, which can provide deeper insight into
  variability and probability—normal or log-normal: that is the question,''
  \emph{BioScience}, vol.~51, no.~5, pp. 341--352, 2001.

\bibitem{kuscu2018modeling}
M.~Kuscu and O.~B. Akan, ``Modeling convection-diffusion-reaction systems for
  microfluidic molecular communications with surface-based receivers in
  internet of bio-nano things,'' \emph{PloS one}, vol.~13, no.~2, p. e0192202,
  2018.

\bibitem{bicen2013system}
A.~O. Bicen and I.~F. Akyildiz, ``System-theoretic analysis and least-squares
  design of microfluidic channels for flow-induced molecular communication,''
  \emph{IEEE Transactions on Signal Processing}, vol.~61, no.~20, pp.
  5000--5013, 2013.

\bibitem{kuscu2016modeling}
M.~Kuscu and O.~B. Akan, ``Modeling and analysis of sinw fet-based molecular
  communication receiver,'' \emph{IEEE Transactions on Communications},
  vol.~64, no.~9, pp. 3708--3721, 2016.

\bibitem{heller2010charge}
I.~Heller, S.~Chatoor, J.~Mannik, M.~A. Zevenbergen, J.~B. Oostinga, A.~F.
  Morpurgo, C.~Dekker, and S.~G. Lemay, ``Charge noise in graphene
  transistors,'' \emph{Nano letters}, vol.~10, no.~5, pp. 1563--1567, 2010.

\bibitem{mucksch2018quantifying}
J.~Mucksch, P.~Blumhardt, M.~T. Strauss, E.~P. Petrov, R.~Jungmann, and
  P.~Schwille, ``Quantifying reversible surface binding via surface-integrated
  fluorescence correlation spectroscopy,'' \emph{Nano Letters}, vol.~18, no.~5,
  pp. 3185--3192, 2018.

\bibitem{vaughan2010bayesian}
S.~Vaughan, ``A bayesian test for periodic signals in red noise,''
  \emph{Monthly Notices of the Royal Astronomical Society}, vol. 402, no.~1,
  pp. 307--320, 2010.

\bibitem{barret2012maximum}
D.~Barret and S.~Vaughan, ``Maximum likelihood fitting of x-ray power density
  spectra: application to high-frequency quasi-periodic oscillations from the
  neutron star x-ray binary 4u1608-522,'' \emph{The Astrophysical Journal},
  vol. 746, no.~2, p. 131, 2012.

\bibitem{sykulski2019debiased}
A.~M. Sykulski, S.~C. Olhede, A.~P. Guillaumin, J.~M. Lilly, and J.~J. Early,
  ``The debiased whittle likelihood,'' \emph{Biometrika}, vol. 106, no.~2, pp.
  251--266, 2019.

\bibitem{anderson1990modeling}
E.~R. Anderson, T.~L. Duvall~Jr, and S.~M. Jefferies, ``Modeling of solar
  oscillation power spectra,'' \emph{The Astrophysical Journal}, vol. 364, pp.
  699--705, 1990.

\bibitem{toutain1994maximum}
T.~Toutain and T.~Appourchaux, ``Maximum likelihood estimators: An application
  to the estimation of the precision of helioseismic measurements,''
  \emph{Astronomy and Astrophysics}, vol. 289, pp. 649--658, 1994.

\bibitem{levin1965power}
M.~Levin, ``Power spectrum parameter estimation,'' \emph{IEEE Transactions on
  Information Theory}, vol.~11, no.~1, pp. 100--107, 1965.

\bibitem{libbrecht1992ultimate}
K.~Libbrecht, ``On the ultimate accuracy of solar oscillation frequency
  measurements,'' \emph{The Astrophysical Journal}, vol. 387, pp. 712--714,
  1992.

\bibitem{smoldyn2022}
S.~S. Andrews, ``Smoldyn: particle-based simulation with rule-based modeling,
  improved molecular interaction and a library interface,''
  \emph{Bioinformatics}, vol.~33, no.~5, pp. 710--717, 2017.

\bibitem{kasdin1995discrete}
N.~J. Kasdin, ``Discrete simulation of colored noise and stochastic processes
  and 1/f/sup/spl alpha//power law noise generation,'' \emph{Proceedings of the
  IEEE}, vol.~83, no.~5, pp. 802--827, 1995.

\bibitem{Optimization}
\BIBentryALTinterwordspacing
T.~M. Inc., ``Optimization toolbox version: 9.2 (r2021b),'' Natick,
  Massachusetts, United States, 2022. [Online]. Available:
  \url{https://www.mathworks.com}
\BIBentrySTDinterwordspacing

\end{thebibliography}

\vfill

\end{document}